%% file: TGF19_TordeuxLebacqueLassarre.tex
\documentclass{svproc}
\usepackage[vmargin=3cm]{geometry}
\usepackage{amsmath,amssymb,tikz}
\newcommand{\dis}{\displaystyle}
\begin{document}
\mainmatter              
\title{Robustness analysis of car-following models for full speed range ACC systems}
\titlerunning{Robustness of car-following models for FACC systems}  
%
\author{Antoine Tordeux\inst{1} \and Jean-Patrick Lebacque\inst{2} \and Sylvain Lassarre\inst{2}}
\authorrunning{A.~Tordeux, J.-P.~Lebacque, S.~Lassarre} 
%
\tocauthor{Antoine Tordeux, Jean-Patrick Lebacque, Sylvain Lassare}
\institute{Division for traffic safety and reliability, University of Wuppertal, Germany\\
{\email{tordeux@uni-wuppertal.de}, ~\texttt{www.vzu.uni-wuppertal.de \vspace{1.5mm}}}
\and GRETTIA, COSYS, IFSTTAR, France\\
{\email{jean-patrick.lebacque@ifsttar.fr}, \email{sylvain.lassarre@ifsttar.fr} \texttt{www.grettia.ifsttar.fr}}
}

\maketitle              

\begin{abstract}
Adaptive cruise control systems are fundamental components of the automation of the driving. 
At upper control level, ACC systems are based on car-following models determining the acceleration rate of a vehicle according to the distance gap to the predecessor and the speed difference. 
The pursuit strategy consists in keeping a constant time gap with the predecessor, as recommended by industrial norms for ACC systems.
The systematic active safety of the systems is tackled thanks to local and string stability analysis.
Several classical constant time gap linear and non-linear car-following models are compared. 
We critically evaluate the stability robustness of the models against latency, noise and measurement error, heterogeneity, or kinetic constraints operating in the dynamics at lower control level. 
The results highlight that many factors can perturb the stability and induce the formation of stop-and-go waves, even for intrinsically stable car-following models.
\end{abstract}

\section{Introduction}
Road vehicles are becoming increasingly automated \cite{VDA2015}. 
Advanced driver assistance systems (ADAS) are common equipment of modern cars, while connected and automated vehicles (CAV) are currently actively developed and tested by manufacturers and suppliers. 
One of the main arguments for the automation of the driving lies in safety aspects. 
Indeed, more than 90\% of traffic accidents are due to human errors \cite{Singh2015}, that could be avoided with safe automatized systems. 
Other arguments for the automation are related to performance, with platooning or optimal traffic assignment, and environment, by limiting jamming and traffic instability (i.e.\ stop-and-go waves), or by facilitating car sharing. 

The automation of road vehicles is commonly classified in 6 levels \cite{SAE2018}. 
Current automation levels are levels 1 and 2 for partial longitudinal and lateral controls in specific driving situations (mainly on highways) and under driver supervision. 
Nowadays, speculations on the levels 3, 4 and 5 of automation without supervision are going well \cite{Litman2018}. 
Experiments in real condition have already shown promising results for partial and full driving automation. 
However, rigorous proofs for the safety of automatized vehicles remain currently actively debated \cite{Warg2014,Koopman2016}. 
Even basic adaptive cruise control (ACC) systems currently available in the market have recently shown during an experiment unsafe behaviors \cite{Gunter2019}.

In ACC systems, the automation is able thanks to sensors, cameras and algorithms, to detect and to track the predecessor and to measure the inter-vehicle distance and relative speed. 
Full speed range ACC systems (FACC) are supposed to regulate the speed even when the vehicle needs to stop. 
ACC and FACC systems are fundamental components of the automation of the driving. 
At upper levels, ACC and FACC systems are based on car-following models determining the acceleration of a vehicle according to the distance ahead and relative speed. 
The challenge consists in finding systematic and parsimonious models able to describe safe and comfortable longitudinal motions. 
The active safety of the models is tackled thanks to local (over-damped) and string (collective) stability analysis (see, e.g., \cite{Darbha1999,Kikuchi2003,Zhou2005,Paden2016}).
Many factors beyond the car-following model may perturb the stability, e.g.\ noise, heterogeneity, delays or again kinetic constraints \cite{Treiber2006,Kesting2008}. 
Safe ACC systems should be robust to those perturbations.

\begin{figure}[!ht]\begin{center}
\input{Figures/scheme_vehicle.tex}\vspace{-1mm}
\caption{Notation used. $x$ is the position of the vehicle and $x_1$ is the position of the predecessor. $D_0\ge0$ is the gap during standstill while  $T_s>0$ is the desired time gap.}\label{fig:scheme}\vspace{-4.5mm}
\end{center}\end{figure}
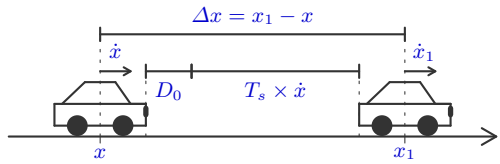

The constant spacing policy, consisting in keeping a constant distance gap independently to the speed, does not describe string stable dynamic \cite{Darbha1999}.
Instead, the current accepted pursuit policy supposes the distance gap proportional to the speed in order to maintain a constant time gap with the predecessor (see \cite{Zhou2005} and Fig.~\ref{fig:scheme}). 
Actual ACC systems proposed by manufacturers offer a choice between pre-defined desired time gap settings, usually ranging from 0.8 to 2.2~s \cite{ISO15622}. 
In this paper, several classical constant time gap linear and non-linear car-following models are compared. 
We calculate the stability conditions and critically evaluate the stability robustness against delay, noise, heterogeneity, or kinetic constraints operating in the dynamics at lower control level. 

\section{Constant time gap models for ACC systems}\label{CFM}
In the following, we consider truncated car-following models with no maximal speed parameter. 
Indeed, our objective is to describe constant time gap interacting pursuit driving, and not a free driving. 
The models are second order ordinary differential equations describing the acceleration rate of a vehicle.

\subsection{Linear car-following models}
The current accepted pursuit policy consists in keeping a constant time gap with the predecessor. 
Indeed, many statistical analyses have shown that a driver tends to keep a constant time gap ranging, roughly speaking, from 1 to 3~s \cite{Banks2003,Tordeux2010}.
The constant time gap strategy is notably the one recommended by the ISO\,15622 norm for ACC systems \cite{ISO15622}. 
Many car-following models allow to maintain a constant time gap with the predecessor. 
The simplest model is a linear version of the optimal velocity (OV) model \cite{Bando1995} 
\begin{equation}
\ddot x(t)=\frac1{T_r}\big[(\Delta x(t)-\ell)/T_s-\dot x(t)\big].
\label{ovm}
\end{equation}
Here, $\Delta x(t)=x_1-x$ is the spacing while $\Delta x(t)-\ell$ is the distance gap, $\ell\ge0$ being the length of the vehicle plus the gap during standstill. 
$T_r>0$ is the relaxation time parameter operating in unsteady state when the time gap is different from the desired time gap $T_s$. 
The OV model is solely based on the distance gap.
A simple linear constant time gap model taking as well in consideration the speed difference term is the full velocity difference model (FVD) \cite{Helly1959,Jiang2001}
\begin{equation}
\ddot x(t)=\frac1{T_r}\big[(\Delta x(t)-\ell)/T_s-\dot x(t)\big]-\frac1{\tilde T_r}\Delta \dot x(t).
\label{fvdm}
\end{equation} 
The constant time gap (CTG) model is a FVD model for which $\tilde T_r=T_s$. 
The CTG model has a single relaxation time parameter $T_r$.
It is frequently used for ACC systems at upper level control (see, e.g., \cite{Zhou2004}).

\subsection{Non-linear car-following models}
OV, FVD and CTG models are linear models that can, even for a queue of vehicles, be explicitly solved. 
We consider in the following two non-linear models that are generally investigated by using numerical simulation. 
The truncated intelligent driver (ID) model \cite{Treiber2000b} with no maximal speed is
\begin{equation}
\ddot x(t)= a \bigg[1 - \bigg(\dot x(t) \frac{T_s-\Delta \dot x(t)\,/\,2\sqrt{ab}}{\Delta x(t)-\ell}\bigg)^2\,\bigg].
\label{idm}
\end{equation}
The model has no relaxation parameters but instead two parameters $a>0$ and $b>0$ for the maximal acceleration and the desired deceleration. 
ID model has been extended to describe ACC systems with adaptive driving strategies \cite{Kesting2000}. 
In the adaptive time gap (ATG) model \cite{Tordeux2010}, the time gap $T(t)=(\Delta x(t)-\ell)/\dot x(t)$ is relaxed to the desired time gap $T_s$, i.e. $\dot T(t)=\lambda\big[T_s-T(t)\big]$. The formulation of the model as an acceleration function is
\begin{equation}
\ddot x(t)=\lambda\dot x(t)\Big[1-\frac{T_s}{T(t)}\Big]+\frac1{T(t)}\Delta \dot x(t).
\label{atg}
\end{equation}
Here $\lambda$ is a relaxation rate parameter. 
The ATG model is close to FVD and CTG models. 
The difference lies in the relaxation times, which are constant parameter in FVD and CTG models, while the relaxation time is the dynamical time gap variable in the ATG model. 
Note that the model is not defined when the speed or distance gap is zero. 
It can however be extended by bounding the time gap variable. 
The bounded ATG model corresponds then to the CTG model.

\section{Linear stability analysis}
Two types of linear stability are classically analysed in the literature: the local stability and the string stability (see, e.g., \cite{Treiber2013}). 
Local stability consists in analysis the dynamics of vehicles following a leader with assigned speed. 
Over-damped conditions specify locally non-oscillating pursuit behavior. 
In string stability, a queue of vehicles, infinite or with periodic boundary conditions, is considered. 

\subsection{Perturbed linear system}
One denotes the car-following model as $\ddot x_n=F(\Delta x_n,\dot x_n,\dot x_{n+1})$, $\Delta x_{n}$ and $\dot x_{n}$ being the spacing and speed of the vehicle $n$, and we suppose the existence of equilibrium spacing and speed solutions $(s,v)$ for which $F(s,v,v)=0$. The dynamics of the perturbations to the equilibrium state $(s,v)$ denoted $\Delta y_n(t)=\Delta x_n(t)-s$ and $\dot y_n(t)=\dot x_n(t)-v$ are linearised to obtain
\begin{equation}
\ddot y_n(t)=a \Delta y_n(t)+b\,\dot y_n(t)+ c\,\dot y_{n+1}(t), 
\label{systlin}
\end{equation}
with $F=F(x,y,z)$, $a=\frac{\partial F}{\partial x}(s,v,v)$, $b=\frac{\partial F}{\partial y}(s,v,v)$ and $c=\frac{\partial F}{\partial z}(s,v,v)$ the model's parameters. 
The uniform solutions $(s,v)$ are linearly stable if $y_n(t)\rightarrow0$ and $v_n(t)\rightarrow0$ as $t\rightarrow0$. 
The predecessor speed $v_{n+1}$ is given in the local stability analysis while it is coupled in the dynamics for the string stability. 
Indeed, string stability condition are more restrictive than the local ones by taking in consideration convective and advective perturbations as well \cite{Wilson2011}. 

\subsection{Stability condition}
Several techniques are used in traffic engineering to analyse the string stability \cite{Orosz2010}. 
Transfer function of the Laplace transform for the perturbation allows to describe the convective or advective nature of the instability. 
String stability condition for a system with periodic boundaries can be obtained by spectrum analysis, while infinite systems are generally tackled using exponential Ansatz $y_n(t)=\Re\big(Ae^{zt+in\theta}\big)$ with $z\in\mathbb C$. 
These two last cases describe the more restrictive stability conditions by considering as well stationary, advective or convective perturbations. 
The stability is hence related in the literature as \emph{absolute string stability} \cite{Wilson2011}. 
The linear stability conditions are established by showing that the solution $z$ of the characteristic equation  
\begin{equation}
z^2 - z (b + c \, e^{i\theta}) + a (1 - e^{i\theta}) = 0
\end{equation}
of the linear differential system (\ref{systlin}) are all defined is the strictly left-half plan of the complex number for all $\theta\in[0,\pi]$ excepted one solution $z_0$ equal to zero for $\theta=0$. 
The conditions for the locations of the zeros of polynomials with complex coefficients have been generalised in \cite[Th.~3.2]{Frank1946}. 
The conditions for the general linear model (\ref{systlin}) are respectively for the local over-damped stability and for the absolute string stability \cite{Wilson2011,Tordeux2012}
\begin{equation}
a>0,~~b^2/4>a\qquad\text{and}\qquad a>0,~~b+|c|<0,~~b^2-c^2>2a.
\label{stab}
\end{equation}
The conditions for the linear and non-linear constant time gap car-following models listed in Sec.~\ref{CFM} are given in Table~\ref{table1}. 
Parameters of OV, FVD and ID models are constraint by the stability conditions, while CTG and ATG models are systematically stable.

\begin{table}
\caption{Local over-damped and string linear stability conditions (\ref{stab}) for OV, FVD, CTG. ID and ATG models with parameters $a,b,c$.}\label{table1}\begin{center}\begin{tabular}{lccc}
\hline\rule{0pt}{12pt}\\[-4mm]
\multicolumn{1}{c}{Model}&$a,b,c$&Local over-damped&Absolute string\\[1mm]
\hline\rule{0pt}{12pt}\\[-4mm]
OV&$\dis\frac1{T_sT_r},\frac{-1}{T_r},0$&$T_s>4T_r$&$T_s>2T_r$\\[2mm]
FVD&$\dis\frac1{T_sT_r},\frac{-1}{T_r}-\frac1{\tilde T_r},\frac1{\tilde T_r}$&$\dis T_s>\frac{4T_r}{(1+T_r/\tilde T_r)^2}$&$\dis T_s>\frac{2T_r\tilde T_r}{2T_r+\tilde T_r}$\\[3mm]
CTG&$\dis\frac1{T_sT_r},\frac{-1}{T_r}-\frac1{T_s},\frac1{T_s}$&$\dis T_s,T_r>0$&$\dis T_s,T_r>0$\\[3mm]
ID&$\dis\frac{2a}{s-\ell},\frac{-2aT_s}{s-\ell}-\frac{\sqrt a}{T\sqrt b},\frac{\sqrt a}{T\sqrt b}$&~~~~~\,$\dis\frac{aT_s^2}{s-\ell}+\sqrt{\frac ab}+\frac{s-\ell}{4bT_s^2}>2$~~~~~&$\dis\frac{aT_s^2}{s-\ell}+\sqrt{\frac ab}>1$\\[3mm]
ATG&$\dis\frac{\lambda}{T_s},\frac{-1}{T_s}-\lambda,\frac1{T_s}$&$\dis T_s,\lambda>0$&$\dis T_s,\lambda>0$\\[3mm]
\hline\rule{0pt}{12pt}\vspace{-10mm}
\end{tabular}\end{center}\end{table}

\section{Stability robustness}
The experiment realised in \cite{Gunter2019} have shown that even recent ACC systems do not describe stable dynamics.
Indeed, many factors operating at lower control level may perturb or even break the linear stability of the car-following planners \cite{Treiber2006,Kesting2008}. 
Example are latency and delay in the dynamics, noise and measurement error, heterogeneity of the behaviors, or even kinetic bound for acceleration or jerk (as recommended by ISO\,15622 norm \cite{ISO15622}). 
The modelling challenge consists in finding models robust against such factors and to establish systematic stabilisation properties. 

\subsection{Latency and response time}
Drivers have reaction time, but they can compensate for this delay by anticipation. 
Automated systems do not react instantaneously as well. 
The perception, computing and control require time and induce computational and mechanical latency in the dynamics. 
For ACC systems, the delay is estimated by 0.5 to 1 second \cite{Zhou2005}. 
The delayed linear car-following model is 
\begin{equation}
\ddot y_n(t+\tau)=a \Delta y_n(t)+b\,\dot y_n(t)+ c\,\dot y_{n+1}(t),\qquad\tau\ge0.
\label{systlind}
\end{equation} 
The corresponding characteristic equation is 
\begin{equation}
\textstyle\lambda^2e^{\lambda\tau} - \lambda (b + c \, e^{i\theta}) + a (1 - e^{i\theta}) = 0,\qquad \theta\in[0,\pi].
\end{equation}
The equation is no more polynomial but exponential-polynomial. 
The generalized Hurwitz conditions do not hold. 
Yet, the location of the zeros can be obtained by subcritical Hopf bifurcation \cite{Orosz2006}, method consisting in investigating purely imaginary zeros and determining the locations by continuity, or by using Taylor approximations for the delayed quantities \cite{Tordeux2012}.
For instance, string stability conditions are $T_s>2T_r$ and $T_s>4\tau(1+\sqrt{1-2T^r/T_s})^{-1}$ for the delayed OV model, while they are $T_s>2\tau$ for CTG and ATG models. The delay in the dynamics clearly results in stability breakdown. 
It can however be (partially) compensated by spatial or temporal anticipation mechanisms \cite{Treiber2006,Tordeux2010}.

\subsection{Stochastic noise}
ACC systems are based on the distance with the predecessor that is measured by sensors such as radar and Fourier signal processing \cite{Reif2010}. 
The radar has a given precision threshold and furthermore factors like the weather perturb the analysis yielding is noise in the measurement. 
This noise may perturb the dynamics and the stability, especially when it depends on the vehicle state \cite{Treiber2006,Hamdar2015}. 
It is for instance reasonable to consider that the precision of the measurement decreases as the distance increases. 
A general stochastic car-following model is $\ddot x=F\big(x_1-x,\dot x,\dot x_1\big)+\xi(\mathbf x,\dot{\mathbf x})$,
where $\xi$ is a noise that may depend on the system state. 
For instance, the CTG model with a Gaussian white noise is   
\begin{equation}
\ddot x(t)=\frac1{T_r}\big[(\Delta x(t)-\ell)/T_s-\dot x(t)\big]-\frac1{T_s}\Delta \dot x(t) +\sigma\dot W(t),
\label{ctgn}
\end{equation}
$\dot W(t)=\text d W(t)/\text dt$ being the derivative of the Wiener process $W$ in the stochastic sense. 
One may write such model as the coupled motion/noise first order models
\begin{equation}
\left\{\begin{array}{lcl}
\dot x(t)&=&\frac1{T_s}[\Delta x(t)-\ell]+\varepsilon(t)\\[1mm]
\dot \varepsilon(t)&=&\frac{-1}{T_r}\varepsilon(t)+\sigma\dot W(t)\end{array}\right.
\label{ou}
\end{equation}
for which the noise $\varepsilon(t)$ is described by the Ornstein-Uhlenbeck process \cite{Uhlenbeck1930}. 
It turns out that the periodic perturbed system with periodic boundary is an Ornstein-Uhlenbeck process as well, whose invariant distribution is a centered Gaussian with explicit covariance matrix \cite{Sato1984,applebaum2015}. 
Indeed, the noise being white in initial CTG, the stochastic linear stability is the deterministic one, i.e.\ stable for all $T_s,T_r>0$. 
No deterministic phase transition is observed, however, noise-induced long-wavelength stop-and-go phenomena are described. 
The stochastic waves can be highlighted by analysing the oscillation in the vehicle speed autocorrelation \cite{Tordeux2016}.  

\subsection{Heterogeneity}
The heterogeneity of the driving behaviors may also play a role in the collective dynamics. 
Indeed, it is observed that low penetration rate of ``stable'' ACC stable allows to homogenise the flows. 
For instance, a single automated vehicle suffices to stabilise 20 vehicles in a ring in \cite{Stern2018}, corresponding to a rate of 5~\%. 
Further simulation experiments corroborate such estimate \cite{Kesting2008b}. 
Yet, if single stable behaviors can stabilise the flow, it is not to exclude that unstable behaviors of certain vehicles may lead to stability breaking. 
The linear perturbed system for heterogeneous vehicles is
\begin{equation}
\frac{\text d}{\text dt}\bar v_n(t)=a_n \,\Delta \bar x_n(t)+b_n \,\bar v_n(t)+ c_n\,\bar v_{n+1}(t),
\end{equation}
where the underlying parameters for the desired time gap, desired speed, vehicle's length etc.\ are specific to each vehicle in a multi-level model.
The corresponding characteristic equation is 
\begin{equation}
\prod_{n=1}^N\big[\lambda^2-b_n\lambda+a_n\big]-e^{i\theta N}\prod_{n=1}^N \big[c_n\lambda+a_n\big]=0,
\end{equation}
while the absolute string stability condition is \cite{Ngoduy2015}
\begin{equation}
\sum_{n=1}^N\frac{b_n^2-c_n^2}{2a_n^2}>\sum_{n=1}^N\frac1{a_n}.
\end{equation}
Note that the condition recovers the classical stability condition $b^2-c^2>2a$ for homogeneous cases.
The non-linear shape of the condition corroborates the fact that single vehicles may strongly influence the collective dynamics. 
Small $a_n$ (i.e.\ large desired time gap) or large $b_n$ (i.e.\ small reaction time) may bring over-weight allowing the condition to hold. 
On the other hand, large $c_n$ (i.e.\ small time gap) for some vehicles lies in stability breaking. 

\subsection{Kinetic constraints}
The ISO\,15622 norm recommends that the acceleration and jerk rates do not exceed safe and comfortable maximal and minimal thresholds. 
The bounds, depending on the speed, are dynamics. 
They range for instance from -5 to -3.5~m/s${^2}$ when the speed increases for the deceleration, when acceleration ranges from -4 to -2~m/s$^2$ \cite{ISO15622}. 
These constraints in the dynamics may be the cause of instability in the models, especially when the relaxation times are small, or even collisions. 
A basic bounded car-following model is 
\begin{equation}
\ddot x(t)=\Big[F\big(x_1(t)-x,\dot x(t),\dot x_1(t)\big)\Big]_{a_{\min}(\dot x(t))}^{a_{\max}(\dot x(t))}
\end{equation}
denoting $[x]^a_b=\min\{a,\max\{b,x\}\}$. 
The acceleration function is continuous. 
It is however not derivable, the partial derivatives do not exist, and the linear stability analysis can not be tackled.
A more regular model can be obtained by using smooth approximation functions, for instance by relaxing the process with anti-windup compensators \cite{Mayr2001,Tarbouriech2010}
\begin{equation}
\dddot x(t)=\frac1{\tau_a}\left(\Big[F\big(x_1(t)-x,\dot x(t),\dot x_1(t)\big)\Big]_{a_{\min}(\dot x(t))}^{a_{\max}(\dot x(t))}-\ddot x_n(t)\right),\quad \tau_a>0.
\end{equation}
Such mechanism induces further inertia into the dynamics that may result in stability breaking. 
Smallness conditions on $\tau_a$ are necessary to ensure stabilisation. 
Further smoothness on the bounding function $[\cdot]_{a_{\min}}^{a_{\max}}$ for the acceleration may also be required.  

\section{Summary}
\begin{figure}[!ht]\begin{center}
\vspace{5mm}
\input{Figures/scheme_summary.tex}\vspace{2mm}
\caption{Operational scheme for ACC systems. At upper level control, ACC systems are stable car-following models based on desired time gap, desired speed or relaxation time parameters. At lower level control, factors such delay and latency, noise and measurement error, heterogeneity and kinetic constraints affect the dynamics and may even break the stability.}\label{fig:summary}\vspace{-3mm}
\end{center}\end{figure}
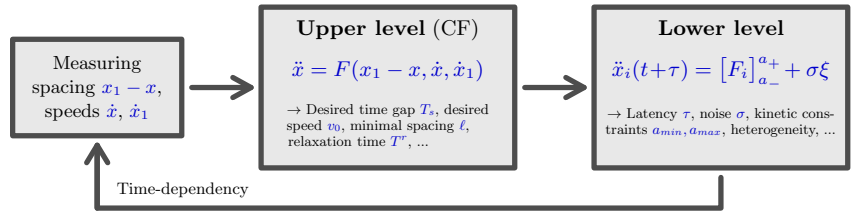

Local and string stability of the car-following models are necessary conditions to demonstrate operational safety properties of ACC systems. 
String stability is in particular fundamental in regard to collective dynamics and the formation of stop-and-go waves that are frequently observed in conventional traffic. 
Many factors operation at lower control level may perturb or even break the stability.
A summary for the architecture of ACC systems and factors affecting the dynamics is proposed in Fig.~\ref{fig:summary}.  
The modelling and safety analysis of ACC and FACC systems consist in determining systematic stability and stabilisation properties of systems of stochastic delayed differential equations under constraints. 
Special cases may be determined analytically (e.g.\ uniform model with periodic boundaries). 
The general problem has many parameter when we consider all factors acting and has to be investigated numerically. 

\end{document}

%% file: Figures/scheme_vehicle.tex
\begin{tikzpicture}[x=.875pt,y=.875pt]
\definecolor{fillColor}{RGB}{255,255,255}
\path[use as bounding box,fill=fillColor,fill opacity=0.00] (0,0) rectangle (227.62, 71.13);
\begin{scope}
\path[clip] (  0.00,  0.00) rectangle (227.62, 71.13);
\definecolor{drawColor}{gray}{0.20}

\path[draw=drawColor,line width= 0.8pt,line join=round,line cap=round] ( 28.19, 16.75) -- ( 67.71, 16.75);

\path[draw=drawColor,line width= 0.8pt,line join=round,line cap=round] ( 28.19, 16.75) -- ( 28.19, 26.16);

\path[draw=drawColor,line width= 0.8pt,line join=round,line cap=round] ( 67.71, 16.75) -- ( 67.71, 26.16);

\path[draw=drawColor,line width= 2.0pt,line join=round,line cap=round] ( 67.71, 21.45) -- ( 67.71, 24.28);

\path[draw=drawColor,line width= 0.8pt,line join=round,line cap=round] ( 28.19, 26.16) -- ( 67.71, 26.16);

\path[draw=drawColor,line width= 0.8pt,line join=round,line cap=round] ( 31.48, 26.16) -- ( 41.36, 35.57);

\path[draw=drawColor,line width= 0.8pt,line join=round,line cap=round] ( 41.36, 35.57) -- ( 54.53, 35.57);

\path[draw=drawColor,line width= 0.8pt,line join=round,line cap=round] ( 54.53, 35.57) -- ( 61.12, 26.16);

\path[draw=drawColor,line width= 0.8pt,line join=round,line cap=round] ( 61.12, 26.16) -- ( 67.71, 26.16);
\definecolor{fillColor}{gray}{0.20}

\path[fill=fillColor] ( 38.07, 16.75) circle (  4.50);

\path[fill=fillColor] ( 57.83, 16.75) circle (  4.50);

\path[draw=drawColor,line width= 0.8pt,line join=round,line cap=round] (159.92, 16.75) -- (199.43, 16.75);

\path[draw=drawColor,line width= 0.8pt,line join=round,line cap=round] (159.92, 16.75) -- (159.92, 26.16);

\path[draw=drawColor,line width= 0.8pt,line join=round,line cap=round] (199.43, 16.75) -- (199.43, 26.16);

\path[draw=drawColor,line width= 2.0pt,line join=round,line cap=round] (199.43, 21.45) -- (199.43, 24.28);

\path[draw=drawColor,line width= 0.8pt,line join=round,line cap=round] (159.92, 26.16) -- (199.43, 26.16);

\path[draw=drawColor,line width= 0.8pt,line join=round,line cap=round] (163.21, 26.16) -- (173.09, 35.57);

\path[draw=drawColor,line width= 0.8pt,line join=round,line cap=round] (173.09, 35.57) -- (186.26, 35.57);

\path[draw=drawColor,line width= 0.8pt,line join=round,line cap=round] (186.26, 35.57) -- (192.85, 26.16);

\path[draw=drawColor,line width= 0.8pt,line join=round,line cap=round] (192.85, 26.16) -- (199.43, 26.16);

\path[fill=fillColor] (169.79, 16.75) circle (  4.50);

\path[fill=fillColor] (189.55, 16.75) circle (  4.50);

\path[draw=drawColor,line width= 0.4pt,line join=round,line cap=round] ( 47.95, 12.04) -- ( 47.95,  9.22);

\path[draw=drawColor,line width= 0.4pt,dash pattern=on 1pt off 3pt ,line join=round,line cap=round] ( 47.95, 12.04) -- ( 47.95, 56.27);

\path[draw=drawColor,line width= 0.4pt,line join=round,line cap=round] (179.67, 12.04) -- (179.67,  9.22);

\path[draw=drawColor,line width= 0.4pt,dash pattern=on 1pt off 3pt ,line join=round,line cap=round] (179.67, 12.04) -- (179.67, 56.27);

\path[draw=drawColor,line width= 0.8pt,line join=round,line cap=round] ( 47.95, 56.27) -- (179.67, 56.27);

\path[draw=drawColor,line width= 0.8pt,line join=round,line cap=round] ( 47.95, 59.16) --
	( 47.95, 56.27) --
	( 47.95, 53.37);

\path[draw=drawColor,line width= 0.8pt,line join=round,line cap=round] (179.67, 53.37) --
	(179.67, 56.27) --
	(179.67, 59.16);

\path[draw=drawColor,line width= 0.4pt,dash pattern=on 1pt off 3pt ,line join=round,line cap=round] ( 67.71, 12.04) -- ( 67.71, 40.27);

\path[draw=drawColor,line width= 0.4pt,dash pattern=on 1pt off 3pt ,line join=round,line cap=round] (159.92, 12.04) -- (159.92, 40.27);

\path[draw=drawColor,line width= 0.8pt,line join=round,line cap=round] ( 67.71, 40.27) -- ( 87.47, 40.27);

\path[draw=drawColor,line width= 0.8pt,line join=round,line cap=round] ( 67.71, 43.16) --
	( 67.71, 40.27) --
	( 67.71, 37.38);

\path[draw=drawColor,line width= 0.8pt,line join=round,line cap=round] ( 87.47, 37.38) --
	( 87.47, 40.27) --
	( 87.47, 43.16);

\path[draw=drawColor,line width= 0.8pt,line join=round,line cap=round] ( 87.47, 40.27) -- (159.92, 40.27);

\path[draw=drawColor,line width= 0.8pt,line join=round,line cap=round] ( 87.47, 43.16) --
	( 87.47, 40.27) --
	( 87.47, 37.38);

\path[draw=drawColor,line width= 0.8pt,line join=round,line cap=round] (159.92, 37.38) --
	(159.92, 40.27) --
	(159.92, 43.16);

\path[draw=drawColor,line width= 0.8pt,line join=round,line cap=round] ( 47.95, 40.27) -- ( 61.12, 40.27);

\path[draw=drawColor,line width= 0.8pt,line join=round,line cap=round] ( 58.12, 38.54) --
	( 61.12, 40.27) --
	( 58.12, 42.00);
\definecolor{drawColor}{RGB}{0,0,0}

\node[text=drawColor,anchor=base,inner sep=0pt, outer sep=0pt, scale=  0.90] at ( 54.53, 45.80) {\textcolor{blue!80!black}{$\dot x$}};
\definecolor{drawColor}{gray}{0.20}

\path[draw=drawColor,line width= 0.8pt,line join=round,line cap=round] (179.67, 40.27) -- (192.85, 40.27);

\path[draw=drawColor,line width= 0.8pt,line join=round,line cap=round] (189.84, 38.54) --
	(192.85, 40.27) --
	(189.84, 42.00);
\definecolor{drawColor}{RGB}{0,0,0}

\node[text=drawColor,anchor=base,inner sep=0pt, outer sep=0pt, scale=  0.90] at (188, 45.80) {\textcolor{blue!80!black}{$\dot x_1$}};

\node[text=drawColor,anchor=base,inner sep=0pt, outer sep=0pt, scale=  0.90] at ( 47.95,  2.68) {\textcolor{blue!80!black}{$x$}};

\node[text=drawColor,anchor=base,inner sep=0pt, outer sep=0pt, scale=  0.90] at (179.67,  3.39) {\textcolor{blue!80!black}{$x_1$}};
\definecolor{drawColor}{gray}{0.20}

\path[draw=drawColor,line width= 0.8pt,line join=round,line cap=round] (  8.43, 12.04) -- (219.19, 12.04);

\path[draw=drawColor,line width= 0.8pt,line join=round,line cap=round] (212.93,  8.43) --
	(219.19, 12.04) --
	(212.93, 15.66);
\definecolor{drawColor}{RGB}{0,0,0}

\node[text=drawColor,anchor=base,inner sep=0pt, outer sep=0pt, scale=  0.90] at (113.81, 61.72) {\textcolor{blue!80!black}{$\Delta x=x_1-x$}};

\node[text=drawColor,anchor=base,inner sep=0pt, outer sep=0pt, scale=  0.90] at ( 77.59, 29) {\textcolor{blue!80!black}{$D_0$}};

\node[text=drawColor,anchor=base,inner sep=0pt, outer sep=0pt, scale=  0.90] at (123.69, 29) {\textcolor{blue!80!black}{$T_s \times\dot x$}};
\end{scope}
\end{tikzpicture}

%% file: Figures/scheme_summary.tex
\begin{tikzpicture}[x=1.1pt,y=1.05pt]
\definecolor{fillColor}{RGB}{255,255,255}
\path[use as bounding box,fill=fillColor,fill opacity=0.00] (0,0) rectangle (310.14, 78.25);
\begin{scope}
\path[clip] (  0.00,  0.00) rectangle (310.14, 78.25);
\definecolor{drawColor}{gray}{0.30}
\definecolor{fillColor}{RGB}{230,230,230}

\path[draw=drawColor,line width= 2.0pt,line join=round,line cap=round,fill=fillColor,fill opacity=0.90] ( 11.49, 63.76) --
	( 69.64, 63.76) --
	( 69.64, 28.98) --
	( 11.49, 28.98) --
	( 11.49, 63.76) --
	cycle;
\definecolor{drawColor}{RGB}{0,0,0}

\node[text=drawColor,anchor=base,inner sep=0pt, outer sep=0pt, scale=  0.80] at ( 40.56, 53.08) {Measuring};

\node[text=drawColor,anchor=base,inner sep=0pt, outer sep=0pt, scale=  0.80] at ( 40.56, 44.30) {spacing \textcolor{blue!80!black}{$x_1-x$},};

\node[text=drawColor,anchor=base,inner sep=0pt, outer sep=0pt, scale=  0.80] at ( 40.56, 35.67) { speeds \textcolor{blue!80!black}{$\dot x$}, \textcolor{blue!80!black}{$\dot x_1$}};
\definecolor{drawColor}{gray}{0.30}

\path[draw=drawColor,line width= 2.0pt,line join=round,line cap=round,fill=fillColor,fill opacity=0.90] ( 96.92, 75.35) --
	(184.14, 75.35) --
	(184.14, 17.39) --
	( 96.92, 17.39) --
	( 96.92, 75.35) --
	cycle;
\definecolor{drawColor}{RGB}{0,0,0}

\node[text=drawColor,anchor=base,inner sep=0pt, outer sep=0pt, scale=  0.90] at (140.53, 65.13) {{\bf Upper level} (CF)};

\node[text=drawColor,anchor=base,inner sep=0pt, outer sep=0pt, scale=  0.90] at (140.53, 50.64) {\textcolor{blue!80!black}{$\ddot x=F(x_1-x,\dot x,\dot x_1)$}};

\node[text=drawColor,anchor=base,inner sep=0pt, outer sep=0pt, scale=  0.60] at (140.53, 36.90) {$\rightarrow$ Desired time gap \textcolor{blue!80!black}{$T_s$}, desired};

\node[text=drawColor,anchor=base,inner sep=0pt, outer sep=0pt, scale=  0.60] at (140.53, 31.10) {speed \textcolor{blue!80!black}{$v_0$}, minimal spacing \textcolor{blue!80!black}{$\ell$},~~~~~};

\node[text=drawColor,anchor=base,inner sep=0pt, outer sep=0pt, scale=  0.60] at (140.53, 25.14) {relaxation time \textcolor{blue!80!black}{$T^r$}, ...~~~~~~~~~~~~~};
\definecolor{drawColor}{gray}{0.30}

\path[draw=drawColor,line width= 2.0pt,line join=round,line cap=round,fill=fillColor,fill opacity=0.90] (211.42, 75.35) --
	(298.65, 75.35) --
	(298.65, 17.39) --
	(211.42, 17.39) --
	(211.42, 75.35) --
	cycle;
\definecolor{drawColor}{RGB}{0,0,0}

\node[text=drawColor,anchor=base,inner sep=0pt, outer sep=0pt, scale=  0.90] at (255.04, 65.13) {{\bf Lower level} };

\node[text=drawColor,anchor=base,inner sep=0pt, outer sep=0pt, scale=  0.90] at (255.04, 50.64) {\textcolor{blue!80!black}{$\ddot x_i(t\!+\!\tau)=\big[F_i\big]_{a_{-}}^{a_{+}}\!+\sigma \xi$}};

\node[text=drawColor,anchor=base,inner sep=0pt, outer sep=0pt, scale=  0.60] at (255.04, 35.45) {$\rightarrow$ Latency \textcolor{blue!80!black}{$\tau$}, noise \textcolor{blue!80!black}{$\sigma$}, kinetic cons-};

\node[text=drawColor,anchor=base,inner sep=0pt, outer sep=0pt, scale=  0.60] at (255.04, 29.65) {traints \textcolor{blue!80!black}{$a_{min},a_{max}$}, heterogeneity, ...};
\definecolor{drawColor}{gray}{0.30}

\path[draw=drawColor,line width= 2.0pt,line join=round,line cap=round] ( 73.08, 46.37) -- ( 93.47, 46.37);

\path[draw=drawColor,line width= 2.0pt,line join=round,line cap=round] ( 87.21, 42.75) --
	( 93.47, 46.37) --
	( 87.21, 49.98);

\path[draw=drawColor,line width= 2.0pt,line join=round,line cap=round] (187.59, 46.37) -- (207.98, 46.37);

\path[draw=drawColor,line width= 2.0pt,line join=round,line cap=round] (201.72, 42.75) --
	(207.98, 46.37) --
	(201.72, 49.98);

\path[draw=drawColor,line width= 2.0pt,line join=round,line cap=round] ( 40.56,  2.90) -- ( 40.56, 25.53);

\path[draw=drawColor,line width= 2.0pt,line join=round,line cap=round] ( 44.18, 19.27) --
	( 40.56, 25.53) --
	( 36.95, 19.27);

\path[draw=drawColor,line width= 2.0pt,line join=round,line cap=round] ( 40.56, 25.53) --
	( 40.56,  2.90) --
	(255.04,  2.90) --
	(255.04, 13.94);
\definecolor{drawColor}{RGB}{0,0,0}

\node[text=drawColor,anchor=base,inner sep=0pt, outer sep=0pt, scale=  0.70] at ( 70,  8) {Time-dependency};
\end{scope}
\end{tikzpicture}

%% file: TGF19_TordeuxLebacqueLassarre.bbl
\begin{thebibliography}{10}

\bibitem{VDA2015}
{Verband der Automobilindustrie e.V.}
\newblock {Automation -- From Driver Assistance Systems to Automated Driving}.
\newblock Technical report, VDA Magazine, 2015.

\bibitem{Singh2015}
S.~Singh.
\newblock Critical reasons for crashes investigated in the national motor
  vehicle crash causation survey.
\newblock Technical report, No. DOT HS 812 115, NHTSA, 2015.

\bibitem{SAE2018}
{SAE International}.
\newblock {Taxonomy and Definitions for Terms Related to On-Road Motor Vehicle
  Automated Driving Systems}.
\newblock Technical report, Standard J3016\_201806, 2018.

\bibitem{Litman2018}
T.~Litman.
\newblock {Autonomous vehicle implementation predictions}.
\newblock Technical report, Victoria Transport Policy Institute, 2018.

\bibitem{Warg2014}
F.~Warg, M.~Gassilewski, J.~Tryggvesson, V.~Izosimov, A.~Werneman, and
  R.~Johansson.
\newblock Defining autonomous functions using iterative hazard analysis and
  requirements refinement.
\newblock In {\em Computer Safety, Reliability, and Security}, pages 286--297.
  Springer International Publishing, 2016.

\bibitem{Koopman2016}
P.~Koopman and M.~Wagner.
\newblock Challenges in autonomous vehicle testing and validation.
\newblock {\em SAE Int. J. Trans. Safety}, 4:15--24, 2016.

\bibitem{Gunter2019}
G.~Gunter, Y.~Wang, D.~Gloudemans, R.~Stern, D.~Work, M.~L.~D. Monache,
  R.~Bhadani, M.~Bunting, R.~Lysecky, J.~Sprinkle, B.~Seibold, and B.~Piccoli.
\newblock String stability of commercial adaptive cruise control vehicles.
\newblock In {\em Proceedings of the 10th ACM/IEEE International Conference on
  Cyber-Physical Systems}, pages 328--329, New York, NY, USA, 2019. ACM.

\bibitem{Darbha1999}
S.~Darbha and K.R. Rajagopal.
\newblock Intelligent cruise control systems and traffic flow stability.
\newblock {\em Transp. Res. C}, 7(6):329--352, 1999.

\bibitem{Kikuchi2003}
S.~Kikuchi, N.~Uno, and M.~Tanaka.
\newblock Impacts of shorter perception-reaction time of adapted cruise
  controlled vehicles on traffic flow and safety.
\newblock {\em J. Transp. Eng.}, 129(2):146--154, 2003.

\bibitem{Zhou2005}
J.~Zhou and H.~Peng.
\newblock Range policy of adaptive cruise control vehicles for improved flow
  stability and string stability.
\newblock {\em IEEE Trans. Intell. Transp. Syst.}, 6(2):229--237, 2005.

\bibitem{Paden2016}
B.~Paden, M.~C{\'a}p, S.~Z. Yong, D.~Yershov, and E.~Frazzoli.
\newblock A survey of motion planning and control techniques for self-driving
  urban vehicles.
\newblock {\em IEEE Trans. Intell. Veh.}, 1(1):33--55, 2016.

\bibitem{Treiber2006}
M.~Treiber, A.~Kesting, and D.~Helbing.
\newblock Delays, inaccuracies and anticipation in microscopic traffic models.
\newblock {\em Physica A}, 360(1):71--88, 2006.

\bibitem{Kesting2008}
A.~Kesting and M.~Treiber.
\newblock How reaction time, update time, and adaptation time influence the
  stability of traffic flow.
\newblock {\em Computer-Aided Civil and Infrastructure Engineering},
  23(2):125--137, 2008.

\bibitem{ISO15622}
{International Organization for Standardization}.
\newblock {Intelligent transport systems -- Adaptive cruise control systems --
  Performance requirements and test procedures}.
\newblock Technical report, Standard ISO 15622:2018, 2018.

\bibitem{Banks2003}
J.~H. Banks.
\newblock Average time gaps in congested freeway flow.
\newblock {\em Transp. Res. A}, 37(6):539--554, 2003.

\bibitem{Tordeux2010}
A.~Tordeux, S.~Lassarre, and M.~Roussignol.
\newblock An adaptive time gap car-following model.
\newblock {\em Transp. Res. B}, 44(8):1115--1131, 2010.

\bibitem{Bando1995}
M.~Bando, K.~Hasebe, A.~Nakayama, A.~Shibata, and Y.~Sugiyama.
\newblock Dynamical model of traffic congestion and numerical simulation.
\newblock {\em Phys. Rev. E}, 51:1035--1042, 1995.

\bibitem{Helly1959}
W.~Helly.
\newblock Simulation of bottlenecks in single lane traffic flow.
\newblock In {\em Proceedings of the Symposium on Theory of Traffic Flow},
  pages 207--238, New York, NY, USA, 1959. Elsevier.

\bibitem{Jiang2001}
R.~Jiang, Q.~Wu, and Z.~Zhu.
\newblock Full velocity difference model for a car-following theory.
\newblock {\em Phys. Rev. E}, 64:017101, 2001.

\bibitem{Zhou2004}
J.~Zhou and H.~Peng.
\newblock String stability conditions of adaptive cruise control algorithms.
\newblock {\em IFAC Proceedings Volumes}, 37(22):649--654, 2004.

\bibitem{Treiber2000b}
M.~Treiber, A.~Hennecke, and D.~Helbing.
\newblock Congested traffic states in empirical observations and microscopic
  simulations.
\newblock {\em Phys. Rev. E}, 62:1805--1824, 2000.

\bibitem{Kesting2000}
A.~Kesting, M.~Treiber, M.~Sch{\"o}nhof, and D.~Helbing.
\newblock Extending adaptive cruise control to adaptive driving strategies.
\newblock {\em Transportation Research Record: Journal of the Transportation
  Research Board}, 2000:16--24, 2007.

\bibitem{Treiber2013}
M.~Treiber and A.~Kesting.
\newblock {\em Traffic Flow Dynamics: Data, Models and Simulation}.
\newblock Springer-Verlag, 2013.

\bibitem{Wilson2011}
R.E. Wilson and J.A. Ward.
\newblock Car-following models: fifty years of linear stability analysis -- a
  mathematical perspective.
\newblock {\em Transport. Plan. Techn.}, 34(1):3--18, 2011.

\bibitem{Orosz2010}
G.~Orosz, R.~E. Wilson, and G.~St{\'e}p{\'a}n.
\newblock Traffic jams: dynamics and control.
\newblock {\em Philosophical Transactions of the Royal Society A: Mathematical,
  Physical and Engineering Sciences}, 368(1928):4455--4479, 2010.

\bibitem{Frank1946}
E.~Frank.
\newblock The location of the zeros of polynomials with complex coefficients.
\newblock {\em Bull. Amer. Math. Soc.}, 52(2):144--157, 1946.

\bibitem{Tordeux2012}
A.~Tordeux, M.~Roussignol, and S.~Lassarre.
\newblock Linear stability analysis of first-order delayed car-following models
  on a ring.
\newblock {\em Phys. Rev. E}, 86:036207, 2012.

\bibitem{Orosz2006}
G.~Orosz and G.~Stepan.
\newblock Subcritical hopf bifurcations in a car-following model with reaction
  time delay.
\newblock {\em Proc. R. Soc. A}, 462(2073):2643--2670, 2006.

\bibitem{Reif2010}
K.~Reif.
\newblock {\em Fahrstabilisierungssysteme und Fahrerassistenzsysteme}.
\newblock Springer, 2010.

\bibitem{Hamdar2015}
S.~H. Hamdar, H.~S. Mahmassani, and M.~Treiber.
\newblock From behavioral psychology to acceleration modeling: Calibration,
  validation, and exploration of drivers’ cognitive and safety parameters in
  a risk-taking environment.
\newblock {\em Transp. Res. B}, 78:32--53, 2015.

\bibitem{Uhlenbeck1930}
G.~E. Uhlenbeck and L.~S. Ornstein.
\newblock On the theory of the brownian motion.
\newblock {\em Phys. Rev.}, 36:823--841, 1930.

\bibitem{Sato1984}
K.-I. Sato and M.~Yamazato.
\newblock Operator-selfdecomposable distributions as limit distributions of
  processes of ornstein-uhlenbeck type.
\newblock {\em Stoch. Proc. Appl.}, 17(1):73--100, 1984.

\bibitem{applebaum2015}
D.~Applebaum.
\newblock Infinite dimensional ornstein-uhlenbeck processes driven by lévy
  processes.
\newblock {\em Probab. Surveys}, 12:33--54, 2015.

\bibitem{Tordeux2016}
A.~Tordeux and A.~Schadschneider.
\newblock White and relaxed noises in optimal velocity models for pedestrian
  flow with stop-and-go waves.
\newblock {\em J. Phys. A}, 49(18):185101, 2016.

\bibitem{Stern2018}
R.~E. Stern, S.~Cui, M.~L.~D. Monache, R.~Bhadani, M.~Bunting, M.~Churchill,
  N.~Hamilton, R.~Haulcy, H.~Pohlmann, F.~Wu, B.~Piccoli, B.~Seibold,
  J.~Sprinkle, and D.~B. Work.
\newblock Dissipation of stop-and-go waves via control of autonomous vehicles:
  Field experiments.
\newblock {\em Transp. Res. C}, 89:205--221, 2018.

\bibitem{Kesting2008b}
A.~Kesting, M.~Treiber, M.~Sch{\"o}nhof, and D.~Helbing.
\newblock Adaptive cruise control design for active congestion avoidance.
\newblock {\em Transp. Res. C}, 16(6):668--683, 2008.

\bibitem{Ngoduy2015}
D.~Ngoduy.
\newblock Effect of the car-following combinations on the instability of
  heterogeneous traffic flow.
\newblock {\em Transportmetrica B}, 3(1):44--58, 2015.

\bibitem{Mayr2001}
R.~Mayr.
\newblock {\em {Regelungsstrategien f{\"u}r die automatische
  Fahrzeugf{\"u}hrung}}.
\newblock Springer-Verlag, 2001.

\bibitem{Tarbouriech2010}
S.~Tarbouriech, G.~Garcia, Gomes da~Silva~Jr., and I.~J.M., Queinnec.
\newblock {\em Stability and Stabilization of Linear Systems with Saturating
  Actuators}.
\newblock Springer-Verlag, 2011.

\end{thebibliography}
